# Benchmark measurements of absolute frequencies and collisional line-shape parameters in the CO–Ar system using comb-based FTS

AKIKO NISHIYAMA[1,2*] GRZEGORZ KOWZAN[1], DOMINIK CHARCZUN[1], ROMAN CIURYŁO[1], PIOTR MASŁOWSKI[1,**]

[1] *Institute of Physics, Faculty of Physics, Astronomy and Informatics, Nicolaus Copernicus University in Toruń, Grudziądzka 5, 87-100 Toruń, Poland*
[2] *National Metrology Institute of Japan (NMIJ), National Institute of Advanced Industrial Science and Technology, 1-1-1 Umezono, 305-8563 Tsukuba, Ibaraki, Japan*
**a-nishiyama@aist.go.jp*
***pima@fizyka.umk.pl*

**Abstract:** We report absolute frequency measurements of the CO fundamental band, together with high-precision Ar-induced collisional line-shape parameters using comb-based Fourier-transform spectroscopy. High-quality spectra were obtained with a stable mid-IR optical frequency comb source and careful suppression of technical noise. By applying analysis methods that fully exploit the advantages of optical frequency comb spectroscopy, precise frequency calibration and robust multi-line fitting were achieved, resulting line-center fitting errors below 0.00001 $cm^{-1}$ for most transitions. The measured transition frequencies agree with HITRAN within 0.0001 $cm^{-1}$, supporting the reliability of the semi-empirical database. Multi-line fits using a speed-dependent Voigt profile yield collisional parameters for approximately 40 P- and R-branch lines, providing a reliable experimental benchmark for CO–Ar line-shape modeling.

## 1. Introduction

High-precision infrared spectroscopy has long provided fundamental insights into molecular energy-level structures and intermolecular interactions, and continues to play an essential role in atmospheric, planetary, and astrophysical research. In particular, accurate characterization of pressure-induced line shifts and line-shape effects is vital for modeling collisional dynamics and for improving the accuracy of spectroscopic databases used to analyze observational data, such as HITRAN [1] and GEISA [2].

Carbon monoxide (CO) is one of the most extensively studied molecules, owing to its significance as a key atmospheric trace gas in both terrestrial and planetary environments [3,4], as well as its role as a prototypical molecule for comparing theory and experiment [5,6]. Nevertheless, the experimental uncertainties associated with collisional shift and broadening parameters in CO–rare-gas perturber systems have not yet been sufficiently reduced especially for the fundamental band. Advanced line-shape analyses have been rather limited [6–11], and comparison with ab-initio calculations that include velocity dependence and Dicke narrowing remains a subject of considerable spectroscopic interest. Moreover, the zero-pressure line positions of the fundamental band currently listed in HITRAN are based on semi-empirical models derived from multiple experimental datasets. The HITRAN line positions differ from the most recent high-resolution Fourier-transform spectroscopy (FTS) measurements [12] by approximately 0.0001 $cm^{-1}$, and are assigned uncertainties in the range of 0.0001-0.001 $cm^{-1}$. Because the accuracy in absolute frequency with conventional FTS is insufficient for a rigorous verification of the HITRAN data due to instrumental line shapes (ILS) or residual phase errors [13–15], their validation and further refinement require the use of frequency-comb-based precision spectroscopy.

The frequency-comb-based FTS overcomes these limitations of conventional FTS by providing broadband spectral coverage together with intrinsically high spectral resolution and high frequency precision [16]. In this approach, the achievable spectral resolution is fundamentally determined by the linewidth of the comb modes themselves, allowing an ILS-free measurement at the ultimate resolution. Furthermore, the high mutual coherence of the comb modes largely suppresses residual phase errors that commonly limit the absolute frequency accuracy in conventional FTS. Nevertheless, achieving ILS-free measurements and high accuracy in line positions with comb-FTS requires careful data analysis. Phase correction to ensure that the interferogram sampling grid matches the comb mode frequencies is essential [17,18]. Absolute frequencies can be determined using simultaneously measured reference lines. Alternatively, they can be inferred from the ILS signature present in the measured spectra, as previously demonstrated in studies of sub-Doppler methane transitions [19]. In the present work, we apply this approach to the CO–Ar system by simultaneously analyzing multiple transitions distributed over the entire branch. In addition, the dominant uncertainty contribution arising from the drift of the He–Ne reference laser is reduced through a multi-line fitting procedure, which enables correction of frequency-scale variations among different measurements.

In this work, we perform measurements of the absolute zero-pressure frequencies and collisional shifts and line shapes of the CO fundamental band (v = 1 ← 0) in CO–Ar mixtures using comb-based FTS. We developed a FTS system based on a stable mid-IR comb that enables acquisition of high SNR broadband spectra. We implemented and validated an analysis algorithm that determines absolute frequencies independently—without relying on auxiliary reference lines—and compared the resulting positions with the semi-empirical values listed in HITRAN. Systematic errors in the line position measurements were mitigated by applying multi-line fitting that fully exploits the advantages of frequency-comb spectroscopy, resulting in reduced uncertainties in the absolute zero-pressure frequencies and the pressure-shift coefficients. We performed multi-pressure fitting of the high-SNR spectra using the advanced line profiles mainly with speed-dependent Voigt profile [20], extracting collisional broadening parameters including the speed-dependent term in a form compatible with the Hartmann–Tran model [21–23] used in HITRAN [24].

## 2. Experiment

An overview of the experimental setup is provided in this section, while further details of the mid-IR frequency-comb source and the overall system can be found in our recent publication [25]. Figure 1 shows a schematic of the mid-IR comb-FTS setup. As the mid-IR comb source, we employ an optical parametric oscillator (OPO) pumped by a mode-locked Yb-fiber laser. The comb wavelength can be tuned from 2.5 to 5 μm by selecting the poling period of the PPLN crystal and its temperature. In this study, we use an output centered at approximately 4.5 μm to probe the CO 0-1 band. The output spectrum has a full width at half maximum (FWHM) of about 80 cm$^{-1}$, which covers more than twenty P- or R-branch transitions within the FWHM bandwidth. The output power at 4.5 μm exceeds 100 mW, which is sufficient for absorption spectroscopy.

The repetition rate ($f_{\mathrm{rep}}$) of the pump laser is 125 MHz and is stabilized to an RF signal. We use the UTC(AOS) frequency standard as the RF reference for all RF signals, providing a relative uncertainty of $2 \times 10^{-13}$ in 1 s and $7.1 \times 10^{-16}$ in a day [26–28]. This repetition rate is directly transferred to the mid-IR comb. The carrier-envelope offset frequency ($f_{\mathrm{ceo}}$) of the mid-IR comb is detected by beating the spectrally broadened pump light with the sum-frequency (pump + idler) light at around 820 nm [29]. The detected $f_{ceo}$ signal is phase-locked to an RF signal by feedback to the OPO cavity length. The resulting relative frequency instability of the mid-IR comb modes is below $10^{-11}$ over the integration period, which is sufficient for the accuracy required in this work.

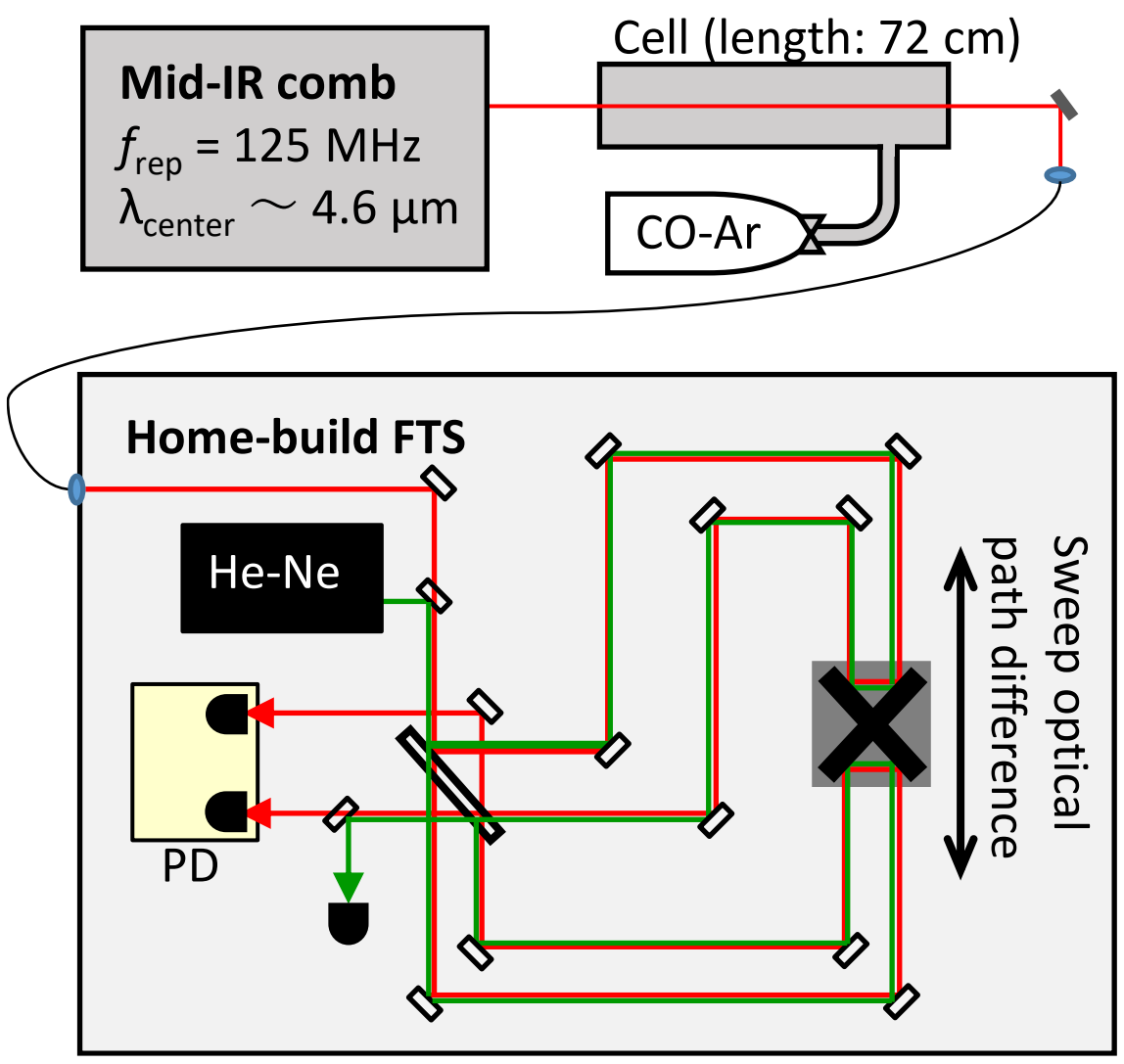

Fig.1. Schematic of the mid-infrared comb-based Fourier-transform spectroscopy (comb-FTS) setup. The mid-IR comb output pass through the spectroscopic cell and coupled to a fiber to send the light to the home-build FTS setup. The mid-IR light share the pass with the He-Ne laser that gives the reference of sampling spacing. We used a balanced MCT detector (PD) for the mid-IR-comb interferogram detection.

The mid-IR comb output is directed into a 72-cm absorption cell filled with sample gas. The transmitted light is coupled into an optical fiber and sent to a home-built FTS system. The interferometer is a basic Mach–Zehnder configuration, but the beams in both arms are retro-reflected on opposite sides of the moving cart to shorten the required mirror-scan distance. An interferogram length of $c/f_{\mathrm{rep}}$(2.4 m) is required for ILS-free measurements. We therefore scanned the moving mirror over 0.6 m. We used an auto-balanced MCT detector (VIGO Photonics, PD) for the mid-IR-comb interferogram detection. A He–Ne laser interferogram was recorded simultaneously in the FTS system and used to define the sampling-point spacing to one-quarter of the He–Ne wavelength.

Measurements were performed at eight pressures ranging from 10 Torr (1.3 kPa) to 400 Torr (53 kPa). The sample consisted of CO diluted in Ar, with the CO concentration adjusted between 0.03% and 0.1% so that the strongest lines exhibited peak absorptions of 70–80%. Because the CO concentration is sufficiently low, collisional effects between CO molecules can be neglected, and the pressure-broadening and shifting can be attributed solely to CO–Ar collisions.

The spectral sampling interval of the comb-based FTS is determined by the comb mode spacing $f_{\mathrm{rep}}$. For low-pressure measurements, where the spectral features are narrower, the spectral point spacing was improved by sweeping the comb-mode frequencies. For example, at a pressure of 10 Torr, an effective resolution of $f_{\mathrm{rep}}/5$ was achieved. Figure 2 shows a normalized spectrum of a CO fundamental (v = 1 ← 0) transition measured at 10 Torr. Each measurement was averaged over 50 interferograms, resulting in an SNR of approximately 1000. To achieve this high SNR, particular care was taken to suppress technical noise, including careful grounding of the detection electronics and optimization of the detector power supply.

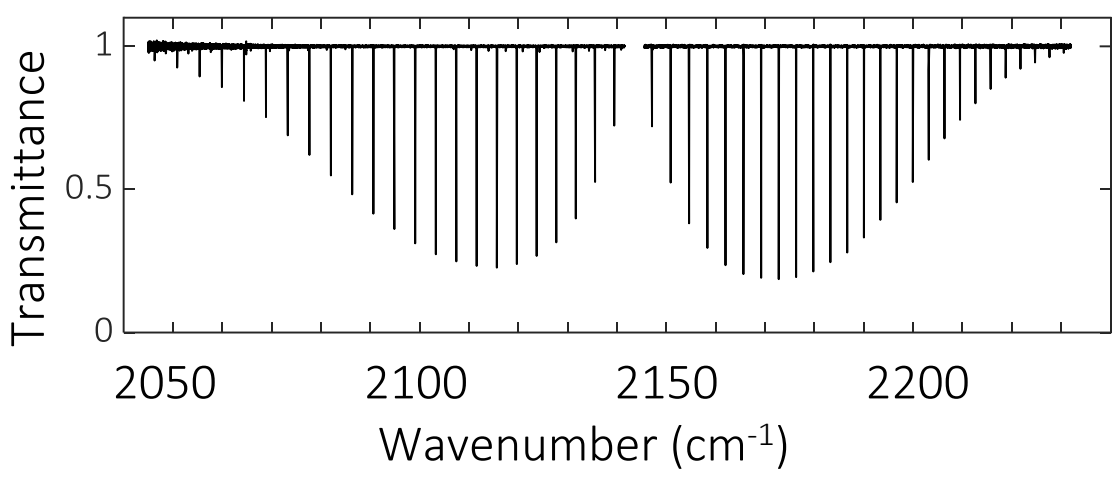

Fig. 2. Normalized CO (v = 1 ← 0) absorption spectrum at 10 Torr. The spectrum represents an average of 50 interferograms and achieves an SNR of ~1000.

As we reported in [25], averaging more than 50 interferograms does not further improve the SNR because the long-term spectral drift of mid-IR comb spectrum becomes dominant. The acquisition time for a single interferogram was 11 s; thus, averaging 50 interferograms required approximately 9 minutes for each frequency setting and pressure condition. Maintaining the spectral stability of the comb source over extended acquisition periods was essential for preserving data quality. For each pressure, we performed an absorption measurement, a background measurement, and a second absorption measurement, yielding two independent spectra per pressure. Including the time required for laser setup and stabilization, the acquisition of data for one CO branch at all eight pressures required approximately two days.

## 3. Results of absolute frequency and shift coefficients

### *3.1 Frequency axis calibration and fitting of the collisional shift*

To suppress the ILS distortions and to provide an accurate frequency axis to the resulting spectrum, a frequency-axis calibration specific to comb-based FTS is required. The first step is the correction of the mismatch between the FTS sampling grid and the optical comb mode frequencies, which accounts for discrepancies between the comb offset frequency and the starting point of the FTS sampling grid, and between $c/f_{rep}$ and the sampling length. Next, the wavelength of the He–Ne laser used as the sampling reference must be calibrated. This is achieved by identifying the He–Ne laser wavelength that minimizes residual ILS distortions through observations of quality-factor (QF) of the fitted line shapes. Here, QF is defined as (absorption depth)/(standard deviation of the fitting residuals). In addition, because the sampling interval is affected by the wavelength dependence of the refractive index of air, this effect is taken into account when determining the absolute frequency scale. Notably, this calibration strategy enables a fully reference-line–free frequency-axis determination, relying solely on the intrinsic properties of the optical frequency comb and the interferometric sampling scheme. Further details of these calibration procedures are provided in the **Supplementary Material**. The procedures to derive He–Ne laser wavelength from QF observation was performed using only 10, 20, and 30 Torr data. For data measured at pressures higher than 50 Torr, the full width at half maximum (FWHM) of the lines exceeds $5 \times f_{rep}$, resulting in a relatively small influence of the ILS and reduced sensitivity for detecting the QF peak. This leads to degraded precision in determining the absolute frequency. Therefore, the QF-based fitting method was not applied to these higher-pressure data. The average value of He–Ne laser wavelength obtained from the low-pressure data acquired on the same day was used for the analysis of high-pressure data.

Figure 3(a) plots the collisional shift frequencies obtained for the R2, R7, and R11 transitions (blue, magenta and green, respectively) from measurements performed at eight different pressures. For the low-pressure data points, the error bars are dominated by the uncertainty of the He–Ne laser wavelength determination and also include the fitting uncertainty of the spectral line center. For measurements at pressures of 50 Torr and higher, the dominant contribution to the error bars arises from the $3\sigma$ fluctuation of the He–Ne laser frequency drift observed over several days of measurements, in addition to the fitting uncertainty of the line center.

Since the speed-dependent term of the CO–Ar collisional shift has been reported to be negligibly small from both theoretical and experimental studies [5,30], it is appropriate to determine the collisional shift coefficient by a linear fit. The result of the linear fit (red line) is also shown in Fig. 3(a). The low-pressure data in the range of 10–30 Torr exhibit a linear pressure dependence, whereas the high-pressure data show larger deviations from the linear fit. The uncertainties of the high-pressure data points caused by He–Ne laser drifts dominate the uncertainties in both the zero-pressure absolute frequency and the collisional-shift coefficient. This interpretation is supported by the observation in all the lines that the data points at each pressure deviate from the linear fit in the same direction and by a similar amount. This behavior indicates that the drift of the He–Ne laser introduces a systematic shift to high pressure measurements.

A key advantage of optical frequency comb spectroscopy is the ability to measure multiple spectral lines simultaneously with high precision. Taking advantage of this feature, we performed a weighted linear fit that simultaneously considered all transitions in order to correct for systematic shifts originating from the He–Ne laser drift that are common to all transitions measured at a given pressure. For each transition $i$ ($i$ = 1, …, $N$) and each pressure point $j$ ($j$ = 1, …, $M$), the measured line center $y_{ij}$ was modeled as

$$y_{ij} = \delta_0^i X_j + \nu_{\text{abs}}^i + \Delta\nu_j,$$

where $\delta_0^i$ is the pressure-shift coefficient of transition $i$, $X_j$ is the pressure of each measurement, $\nu_{\text{abs}}^i$ is the zero-pressure extrapolated value (i.e., the zero-pressure line position). The term $\Delta\nu_j$ represents a systematic shift component common to all spectral lines at pressure point $j$. The parameters $\delta_0^i$ and $\nu_{\text{abs}}^i$ were determined using a weighted least-squares method, where the uncertainties $\sigma_{ij}$ of the line-center frequencies obtained from the spectral fitting were used as weights. For the low-pressure data in the range of 10–30 Torr, for which the He–Ne laser wavelength was determined by the QF observations, the influence of the He–Ne laser drift can be neglected; therefore, $\Delta\nu_j$ was fixed to zero for these data points. The standard uncertainties of the fitted parameters were calculated from the covariance matrix. Using the obtained values of $\Delta\nu_j$, the measured data were corrected for the He–Ne laser drift-induced systematic shift, and the pressure-shift coefficients and zero-pressure line positions for each transition were derived.

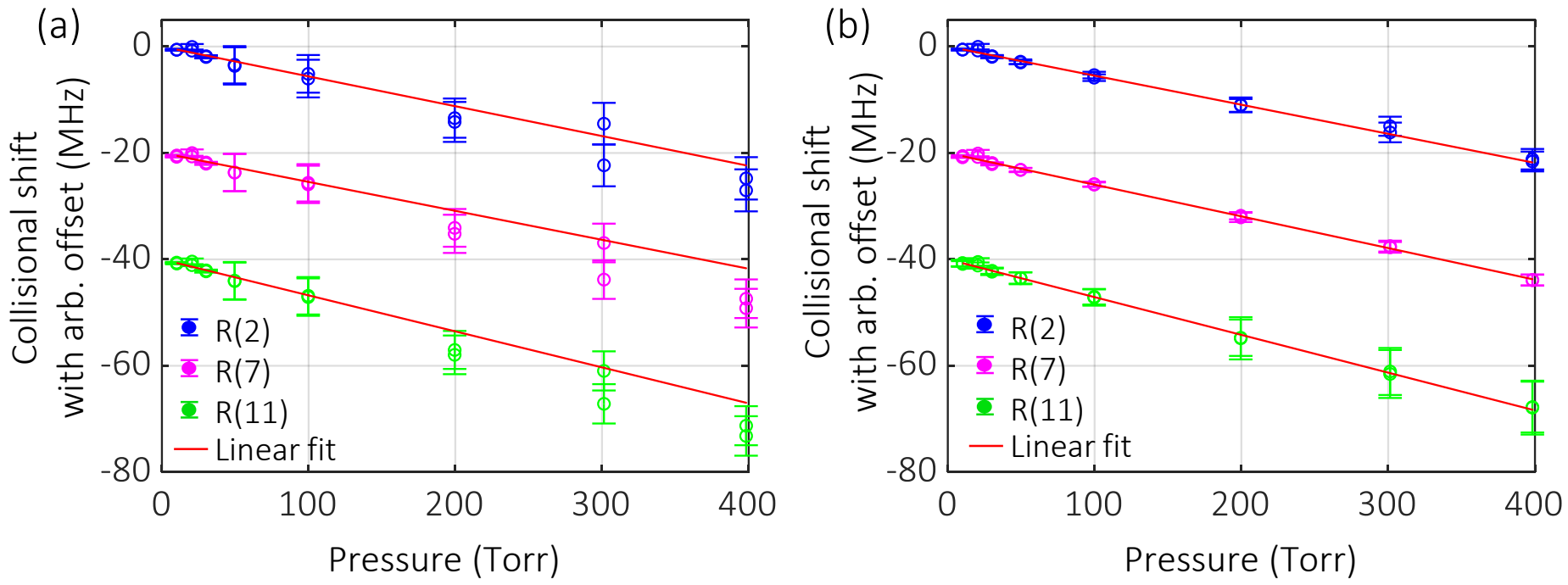

Fig. 3(a). Pressure-induced frequency shifts of the R2, R7, and R11 transitions (blue, magenta, and green, respectively) measured at eight pressures and linear fits (red lines). For clarity, an arbitrary vertical offset is applied to each transition to display only the relative pressure-dependent frequency shifts. The error bars represent the uncertainties of the line-center frequencies, including the contributions from the spectral fitting and the He–Ne laser frequency uncertainty or drift. (b)Drift-corrected transition frequencies of the R2, R7, and R11 transitions (blue, magenta, and green, respectively) and the corresponding linear fits (red lines). Compared with the uncorrected data presented in Fig. 3(a), the high-pressure data exhibit significantly improved agreement with the linear fits after correction for the He–Ne laser frequency drift. The error bars represent only the uncertainties from the spectral fitting.

Figure 3(b) shows the drift-corrected transition frequencies of the R2, R7, and R11 transitions (blue, magenta and green, respectively) together with the corresponding linear fits (red line). Compared with the uncorrected results shown in Fig. 3(a), the high-pressure data exhibit significantly improved agreement with the linear fit. Similar results were obtained for approximately 20 lines in each of the P and R branches.

### *3.2 Determination of the zero-pressure line positions*

Figure 4 shows the absolute zero-pressure transition frequencies, presented as differences from the values reported in the HITRAN database. The horizontal axis is given by the line number $m$, defined as $m = -J''$ for the P branch and $m = J'' + 1$ for the R branch. The results are plotted using different colors to distinguish those derived from fits using only the low-pressure data (green circle) and global fit of all transitions including a drift-correction term (blue circle). The error bars represent the uncertainty of the linear fit. The latest FTIR results by Devi *et al.* [12] are also shown in the same plot (red triangle). When only the low-pressure data are used the resulting uncertainties are relatively large. By contrast, the inclusion of the high-pressure data with the drift-correction term significantly reduces the uncertainty in the determination of the absolute transition frequencies. The resulting linear-fit uncertainties are below 0.00001 $cm^{-1}$ for most of the measured transitions. In the drift-corrected results, the differences from the HITRAN values are within a few hundred kilohertz for lines with high signal-to-noise ratio. These results show improved agreement with HITRAN compared to the most recent FTS results reported by Devi *et al*. [12], and the observed deviations are smaller than the uncertainties quoted in the HITRAN database. A table listing the absolute zero-pressure transition frequencies for all measured lines is given in the Supplementary Material (Table s1).

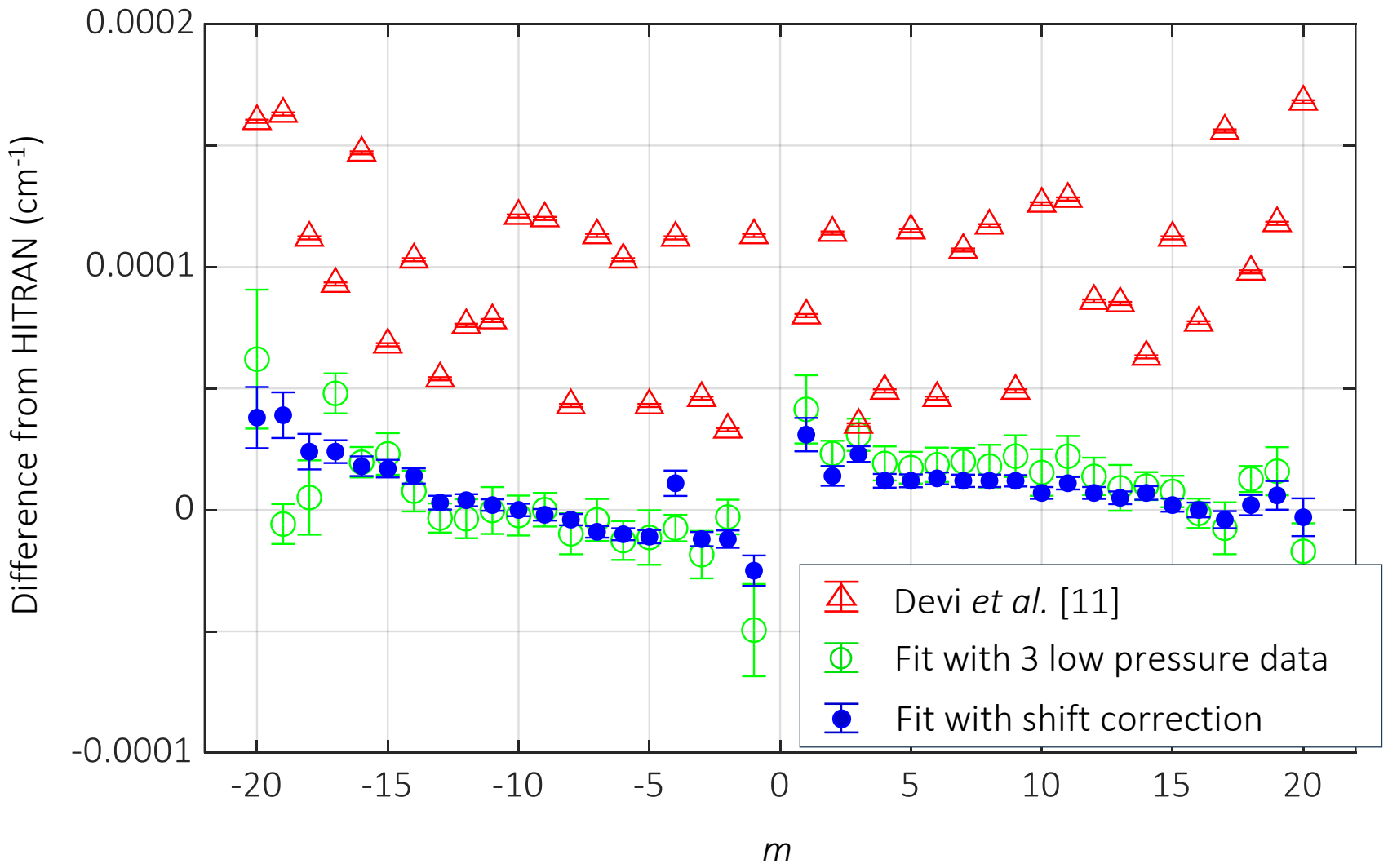

Fig. 4. Absolute zero-pressure line positions expressed as differences from the HITRAN database. Results obtained from linear fits using only the low-pressure data are shown as green circles, while those derived from a global fit including all pressure data and the drift-correction term are shown as blue circles. Error bars represent the uncertainties of the linear fits. For comparison, recent FTIR measurements reported by Devi *et al.* [11] are also included (red triangles). A slight anomaly in the m dependence at m = −4 is likely caused by the overlap with an isotopic transition.

Moreover, the deviations appear to be systematic and wavelength-dependent rather than randomly distributed. This behavior is of particular interest for assessing the uncertainty of the absolute frequency measurements with our comb-FTS system. HITRAN reports the absolute zero-pressure transition frequencies with a stated uncertainty of 0.0001 to 0.001 $cm^{-1}$, and these values are derived using semi-empirical methods. Since the agreement achieved in the present work is already better than the uncertainty reported in HITRAN, it is not possible to unambiguously determine whether the remaining discrepancies originate from the HITRAN values or from the present measurements. The observed deviations might reflect imperfections in the air-refractive-index correction used in the present analysis. The refractive index of air described by the Mather formula requires corrections for temperature, pressure, and humidity. In the present experiment only the laboratory temperature was monitored, and no corrections were applied for pressure or humidity. This limitation can be completely eliminated by employing an FTS under vacuum, which will further improve the accuracy of comb-FTS system.

### *3.3 Collisional shift parameters*

Figure 5 plots the pressure-shift coefficients $\delta_0$ obtained simultaneously with the zero-shift line positions from the linear fits of the line-center frequencies. The error bars represent the uncertainty in the slope of the linear fit. As in the case of the absolute frequency determination, fits based only on the low-pressure data exhibit relatively large uncertainties. In contrast, the fits including the drift-correction term yield significantly smaller uncertainties, allowing a clear dependence on the rotational quantum number *m* to be resolved. A table listing the pressure-shift coefficients $\delta_0$ for all measured lines is given in the Supplementary Material (Table s2).

Pressure-shift coefficients reported in the literatures [5,30] are also shown in the same plot for comparison. The results reported by Luo *et al* [30] were obtained using continuous-wave (CW) laser spectroscopy with difference frequency generation (DFG) [7,31,32], while those by Kowzan *et al.* [5] were derived from *ab initio* calculations. The average cell temperature during the present measurements was 299.2 K. The measurements reported in Luo *et al.* [30] were performed at 296 K. For Kowzan *et al*. [5], the temperature dependence of the parameters was provided; therefore, reported values were corrected to 299.2 K for comparison. Compared with the results of Luo *et al.* [30], most of the measured lines agree within the uncertainties.

Previous high-precision pressure-shift measurements [30] were performed using CW infrared radiation generated by DFG of two visible laser beams overlapped in a nonlinear crystal. In these experiments, the infrared radiation was split into three beams to simultaneously monitor the IR intensity, the absorption in a low-pressure reference cell, and the absorption in the measurement cell, where the gas pressure and temperature could be varied. The present measurements exhibit small uncertainties that are comparable to those obtained in previous CW-laser-based measurements. This performance benefits from the advantages of the comb spectroscopy, which enables broadband spectral acquisition within a short measurement time, facilitates efficient measurements at multiple pressures, and allows the simultaneous analysis of multiple transitions. These capabilities make it possible to mitigate systematic errors through global fitting. Furthermore, the observed smooth dependence of the pressure-shift parameters on the line number $m$ provides additional evidence that the experimental uncertainties are very small.

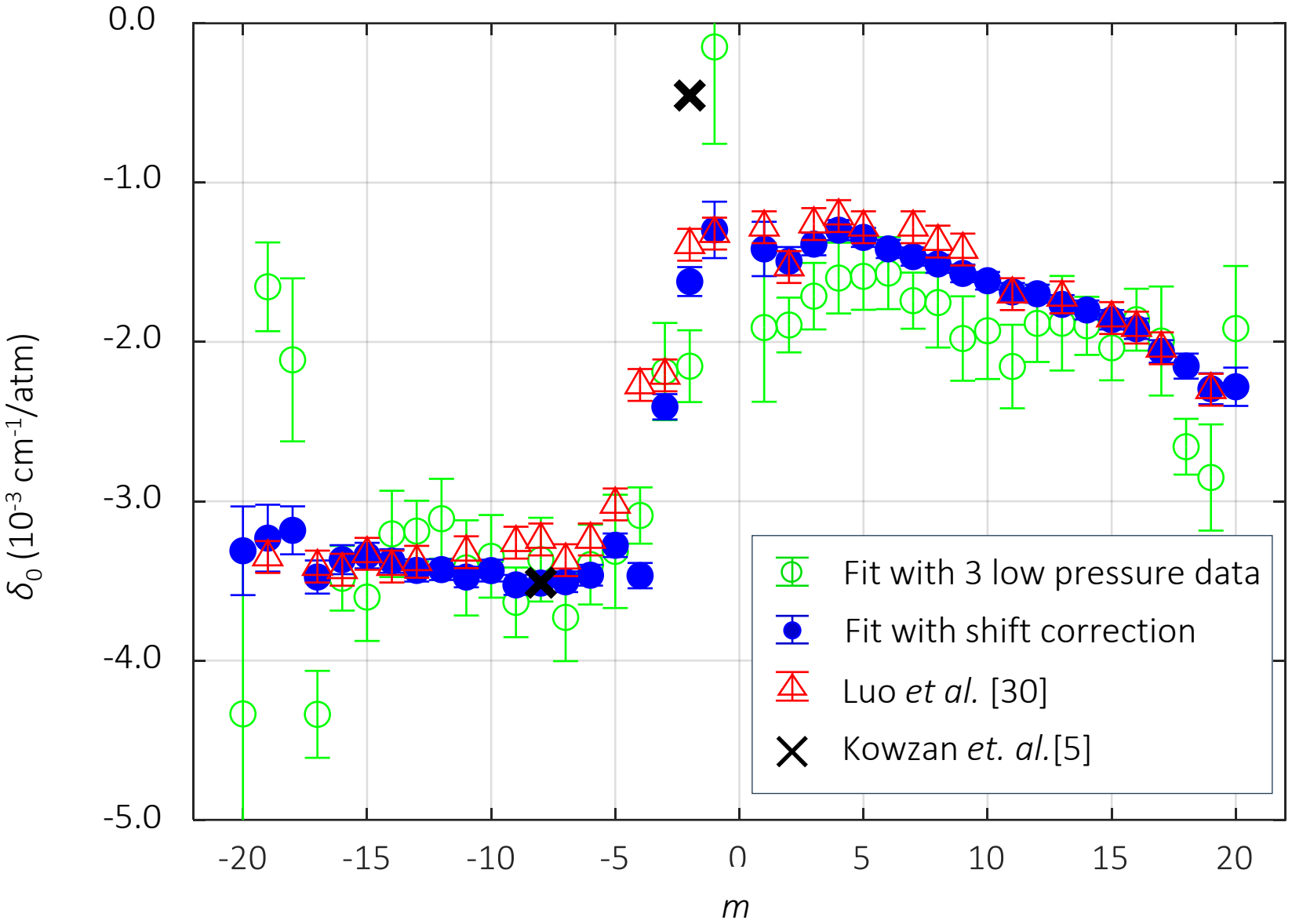

Fig. 5. Pressure-shift coefficients $\delta_0$obtained simultaneously with the absolute transition frequencies from linear fits to the measured line-center frequencies, plotted as a function of the $m$. Error bars represent the uncertainties of the fitted slopes. Results derived from fits using only the low-pressure data are shown by green circles and those from global fits including a drift-correction term are shown by blue circles. Literature values from CW-laser measurements Luo *et al* [30](red triangle) and *ab initio* calculations by Kowzan *et al* [5](black cross) are included for comparison. A slight anomaly in the $m$ dependence at m = −4 is likely caused by the overlap with an isotopic transition.

## 4. Fitting with collisional effects with advanced line shapes

### *4.1 Short description of the line-shape models*

Accurate analysis of molecular absorption spectra critically depends on the choice of an appropriate line-shape model that properly describes both Doppler and collisional broadening mechanisms. The simplest Voigt profile (VP) represents these two effects as a convolution; however, real molecular collisions involve additional effects, such as speed-dependent broadening and frequency shifts [20], as well as Dicke narrowing caused by velocity-changing collisions [33]. To account for such effects, many different models were proposed over the years [34–36]. More recently, based on line shapes proposed in Refs. [21,22,37,38], the Hartmann-Tran profile (HTP) [23] has been widely adopted. HTP is the partially-corelated quadratic-speed-dependent hard-collision profile, whose parametrization enables accurate reproduction of molecular line shapes while maintaining reasonable computational efficiency [24,39].

For the CO–Ar system, Wehr *et al* [9] performed an analysis based on a quantum Boltzmann equation (the generalized Waldmann-Snider equation) with a hard-sphere collision kernel modeling velocity-changing collisions, and demonstrated that the actual magnitude of Dicke narrowing is 70–90% smaller than that predicted by mass diffusion models. This result is consistent with theoretical predictions that rotationally inelastic collisions do not contribute to Dicke narrowing for isolated lines, highlighting the necessity of physically consistent collision models rather than purely empirical descriptions in line-shape analyses of CO–Ar spectra. More recently, ab initio molecular collision calculations have been used to derive the full set of parameters of the Hartmann–Tran profile (HTP) and have been successfully applied to the CO–Ar system [5]. In that work, pressure broadening coefficients were calculated using high-accuracy interaction potentials.

Based on these considerations, we primarily employ an HTP-compatible speed-dependent Voigt profile (SDVP) [20] for fitting of the measured spectra, and compare the resulting line-shape parameters with recent experimental measurements and theoretical predictions.

### *4.2. Results of multi-line fit and collisional broadening parameters*

To suppress residual ILS distortions caused by the non-flatness of the interferogram, we applied an amplitude correction prior to Fourier transformation. The correction procedure is described in our previous work [25] and in the **Supplementary Material**. Figure 6 shows the measured line shapes of the R7 transition recorded over the pressure range of 10–400 Torr, together with the residuals obtained from single-pressure fits using the VP (magenta), and from both single-pressure and multi-pressure fits using the SDVP (green and blue, respectively). In the single-pressure fits, the line-center frequency, line intensity, pressure-broadening parameter $\gamma_0$, speed-dependence parameter $\alpha$ for SDVP, and baseline were treated as independent fit parameters, while the Doppler width was fixed to the value calculated from the measured cell temperature. In the multi-pressure fits, the $\gamma_0$ and $\alpha$ were constrained to be common across all pressures.

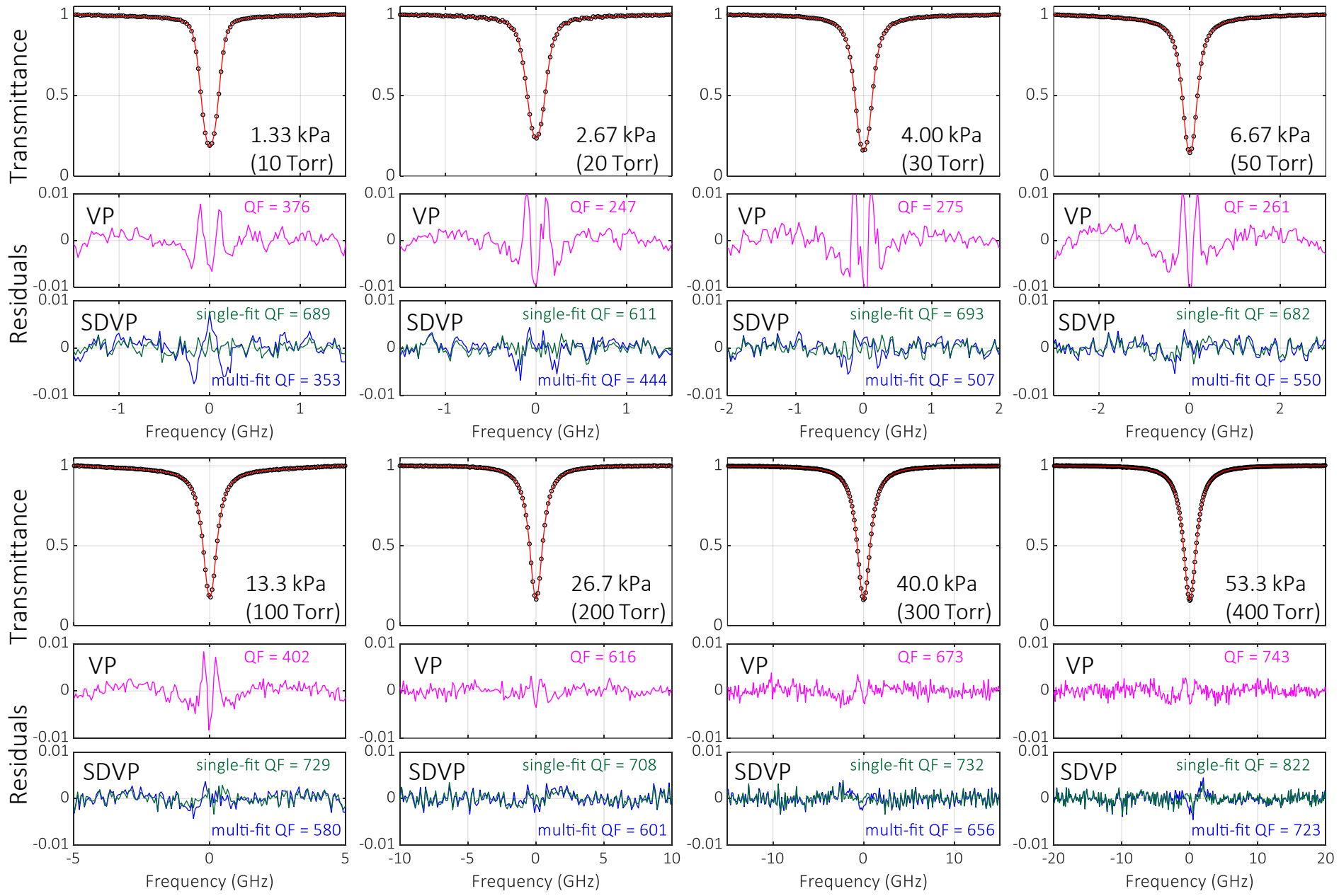

Fig. 6: Measured line shapes of the CO-Ar R7 transition recorded at pressures from 10 to 400 Torr. The residuals obtained from single-pressure fits using the Voigt profile (VP, magenta), single-pressure fits using the speed-dependent Voigt profile (SDVP, green), and multi-line SDVP fits (blue) are shown below each spectrum.

The VP fits exhibit pronounced residuals, particularly in the pressure range of 10–100 Torr, with a characteristic "w-shaped" pattern. This behavior indicates that the present measurements realized the high accuracy beyond the Voigt model limit. By introducing the speed-dependent term through the SDVP, these residuals are effectively eliminated, and the quality factor improves by more than 300 at low pressures compared with the VP fits.

A comparison between the single-pressure and multi-pressure SDVP fits reveals that the residuals of the multi-pressure fit become slightly larger at low pressures. The spectra reconstructed from the fitted parameters exhibit marginally narrower wings than the measured spectra in this pressure regime. This trend is consistently observed for all P- and R-branch lines. At this level of residuals, a contribution from residual instrumental line shape (ILS) effects in the comb-based FTS measurements cannot be excluded. However, if the residual ILS can be neglected, these results suggest that the adopted line-shape model remains incomplete, and that additional parameters are required to fully describe the collisional line profile.

Multi-pressure fits [40,41] using the speed-dependent Nelkin–Ghatak profile [21,22] were also attempted for several transitions, but stable convergence could not be achieved due to the correlation between the Dicke narrowing parameter $\upsilon_{opt}$ and the speed-dependent term $\alpha$. When $\upsilon_{opt}$ was fixed to the value calculated in ref [5], the resulting multi-pressure fits exhibited significantly larger residuals, indicating that $\upsilon_{opt}$ is nearly an order of magnitude smaller than the ab initio value reported in ref [5]. These results confirm observations made for transitions in the second overtone band [42].

Figure 7(a) shows the pressure broadening coefficient $\gamma_0$ obtained from the SDVP multi-pressure fits, plotted as a function of $m$. The error bars represent the uncertainties of the fitted parameters and are smaller than the symbol sizes. For comparison, results from cw-laser measurements Luo *et al* [30] (red circle) and Wehr *et al* [9] (cyan triangle), and theoretical calculations by Kowzan *et al* [5] (black cross) are also included in the same plot. The values reported in ref [30] were obtained using VP fitting at 296 K, those from ref [9] were derived using the SDBBP model and have been corrected to 299.2 K. Compared with ref [30], the present results clearly yield systematically larger values of $\gamma_0$. The differences between the previously reported values and the present results are shown in Fig. 7(b). The deviation from ref [30] exhibits a clear $m$ dependence that closely resembles the $m$ dependence observed for the speed-dependence parameter $a$ which is presented in Fig. 8. This similarity indicates that the discrepancy mainly originates from differences in the adopted line-shape models; another possible reason is the temperature difference between our measurements and Ref. [30], as discussed in Ref. [25]. Both the Wehr *et al* [9] and Kowzan *et al* [5] results are approximately 1% smaller than the present values of $\gamma_0$. CO-Ar lines are narrowed by speed dependence of pressure broadening and by velocity-changing collisions. The Voigt profile does not explicitly model these effects, therefore it fits narrowed lines by underestimating pressure broadening cross sections. On the other hand, the SDVP overestimates narrowing due to speed dependence of broadening, yielding larger thermally-averaged values. The SDBBP fits and ab initio calculations accurately model both mechanisms, which puts their $\gamma_0$ values between those of VP and SDVP. The dependence of $\gamma_0$ on the adopted line-shape model has also been discussed in Ref. [25].

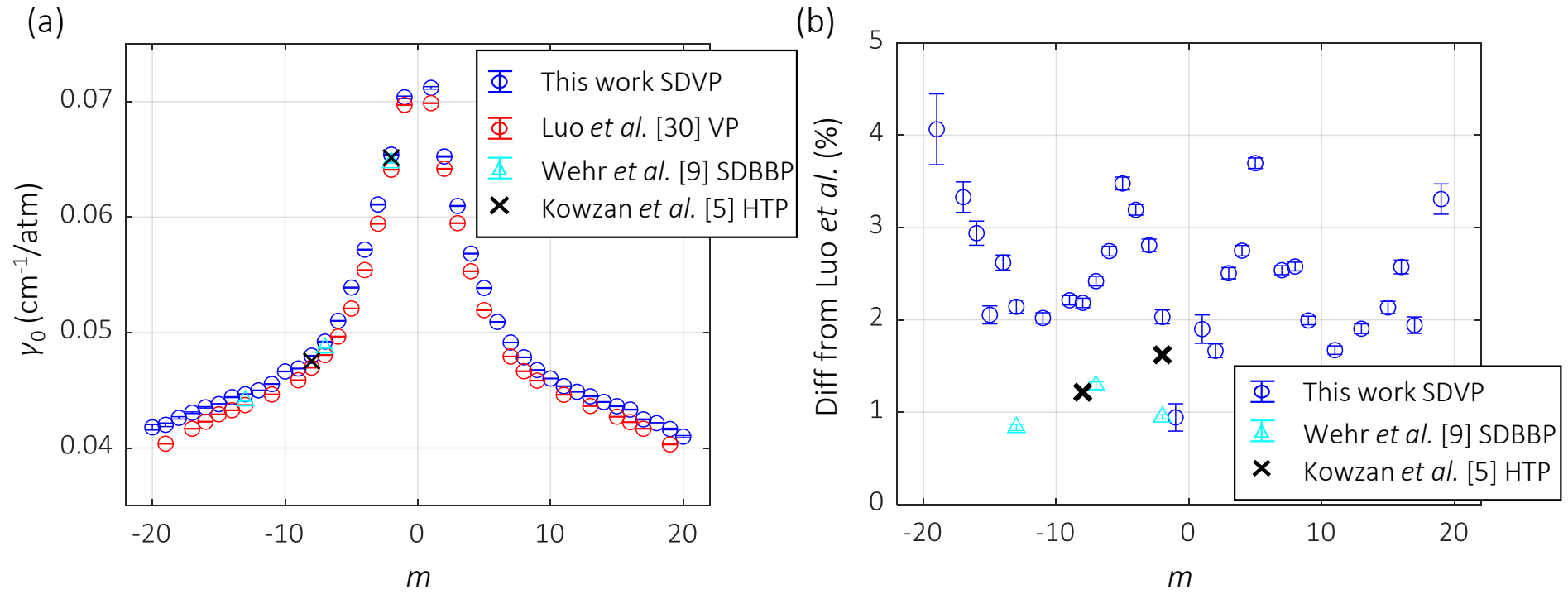

Fig. 7. (a) Pressure broadening coefficient $\gamma_0$ obtained from the SDVP multi-pressure fits as a function of the rotational quantum number $m$ (blue circles, measured at 299.2 K). The error bars represent the uncertainties of the fitted parameters. For comparison, results from cw-laser measurements by Luo *et al*. [30] (red circles, measured at 296 K) and Wehr *et al*. [9] (cyan triangles, measured at 294 K and corrected to 299.2 K), as well as theoretical calculations by Kowzan *et al* [5] (black crosses, corrected to 299.2 K), are also shown. (b) Differences between the pressure broadening coefficients reported in Luo *et al*. [30] and the present results as a function of $m$. The observed $m$ dependence closely resembles that of the speed-dependence parameter $a$, which is shown in Fig. 8.

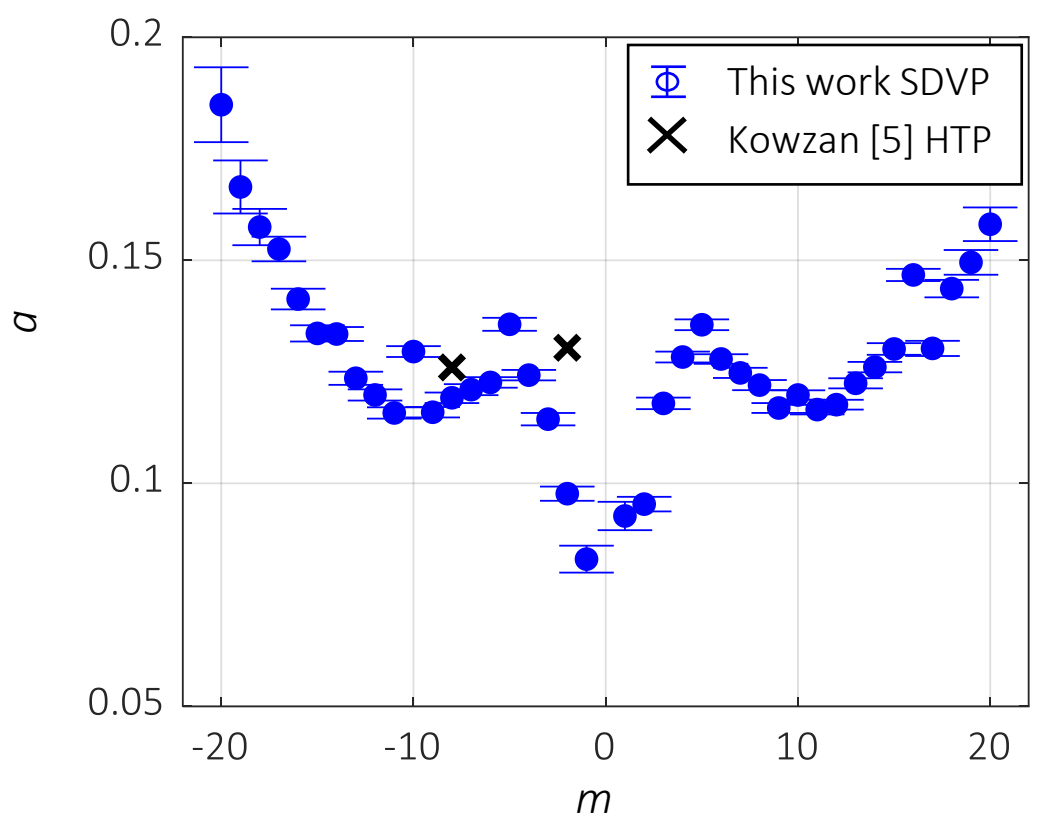

**Fig. 8.** Speed-dependence parameters $a$ obtained from the SDVP multi-pressure fits, plotted as a function of the line number $m$ (blue). The error bars represent the uncertainties of the fitted parameters and are smaller than the symbol sizes. Theoretical values calculated by Kowzan *et al* [5] are also shown for comparison (black crosses).

Next, the values of $a$ obtained from the multi-pressure fits are plotted as a function of $m$ in Fig. 8. The multi-pressure fitting approach significantly reduces the uncertainty in $a$, allowing the $m$ dependence to be clearly resolved. The distribution is symmetric with respect to positive and negative $m$, exhibiting a minimum around $m = 0$ and a maximum near $m \approx 5$, followed by a parabolic increase at higher $|m|$ values. Also shown in Fig. 8 are the theoretical results reported by Kowzan *et al.* [5] (black crosses). The differences between the present experimental values and the Kowzan *et al* predictions are relatively small, indicating overall good agreement. A table listing the values of $\gamma_0$ and $a$ for all measured lines is given in the Supplementary Material (Table s2).

## 5. Summary

In this work, we demonstrated a methodology for high-accuracy absolute frequency determination using comb-based FTS over a wide pressure range. To fully exploit the broadband capability of optical frequency comb spectroscopy, we further introduced a global fitting approach that enables efficient mitigation of systematic errors across multiple transitions and pressures. By applying a series of systematic corrections, we reported zero-pressure line positions for 40 lines in the fundamental band of CO with the highest experimental accuracy achieved to date, with fitting errors below 0.00001$cm^{-1}$ for most of the measured lines. The resulting line positions show excellent agreement with the HITRAN database with deviations smaller than the uncertainties quoted in HITRAN, thereby providing strong experimental support for the semi-empirical methods underlying the database. The collisional shift coefficients were also determined with high precision over the entire band with smaller uncertainties than the previous CW laser measurements, establishing a new benchmark dataset for pressure-shift parameters in CO–Ar. In addition, high-precision line-shape parameters were derived using the SDVP, which are compatible with the HTP. Using this approach, we report, for the first time, speed-dependent broadening parameters $a$ for multiple transitions within the fundamental vibrational band of CO–Ar. These results constitute a stringent experimental benchmark for future theoretical studies and are expected to contribute to the development and validation of more rigorous *ab initio* models of collisional effects in molecular spectroscopy.

## Funding

This research was supported by National Science Centre, Poland project no. 2019/35/D/ST2/04114. The research is part of the program of the National Laboratory FAMO in Toruń, Poland. AN acknowledges the support from the Yamada Science foundation Overseas research support 2024. GK acknowledges the support from the European Union's Horizon 2020 Research and Innovation Program under Marie Sklodowska-Curie Grant Agreement No 101028278.

## Acknowledgement

We thank Dr hab. inż. Grzegorz Soboń (Wroclaw University of Science and Technology) for useful discussions on the development of the mid-IR comb system and for providing the PCF fiber for the OPO setup.

# Benchmark measurements of absolute frequencies and collisional line-shape parameters in the CO–Ar system using comb-based FTS: supplemental document

### *Analytical protocol for absolute frequency determination*

As detailed in Sec. 2.3 of Rutkowski *et al.*[17] , comb-based FTS requires that the FTS sampling frequencies be matched to the comb mode frequencies. This requirement becomes particularly critical when the spectral linewidths are comparable to the comb $f_{\text{rep}}$. The mismatch between the sampling grid and the comb mode frequencies causes instrumental line shape (ILS) in the retrieved spectrum. The first source of mismatch arises from the fact that the FTS sampling grid begins at zero frequency, whereas the comb modes have offset by the carrier-envelope offset frequency $f_{\text{ceo}}$. This discrepancy can be corrected by applying a frequency shift equal to the measured $f_{\text{ceo}}$ to the FTS-derived spectrum. In practice, this is accomplished by multiplying the interferogram $P(\Delta)$ by an exponential term containing $f_{\text{ceo}}$ prior to performing the fast Fourier transform (FFT).

$$S_{FTS}(\nu_{FTS}) = Aabs\{FFT[P(\Delta)\exp(-i2\pi f_{ceo}\Delta/c)]\} \quad \text{(s1)}$$

Where $A$ is an overall amplitude scaling factor, and $c$ is the speed of light. During the measurement, the carrier–envelope offset frequency $f_{ceo}$ must remain sufficiently stable.

The second source of mismatch arises from the discrepancy between required interferogram length and the actual recorded interferogram length. To realize ILS-free measurement, the interferogram length must be truncated to exactly $c/f_{\text{rep}}$, corresponding to the spacing between adjacent comb pulses. The sampling interval, $x_{\text{sampling}}$, is approximately 633/4 nm, which corresponds to a relative resolution of only $10^{-8}$with respect to the interferogram length $c/f_{\text{rep}}$. The difference between the FTS sampling frequency ($f_{\text{FTS}}$) and the $f_{\text{rep}}$ is written as,

$$\delta = f_{\text{FTS}} - f_{\text{rep}} = \frac{c}{x_{\text{sampling}}\,k} - f_{\text{rep}}. \quad \text{(s2)}$$

here, $k$ is the number of data points closest to $c/f_{\text{rep}}$. The resulting mismatch in the comb-mode frequencies, using the comb mode index $M$ at the spectral center of the mid-infrared comb, becomes

$$\nu_{\text{shift}} = -M\,\delta. \quad \text{(s3)}$$

This frequency shift is corrected, analogously to the $f_{\text{ceo}}$ correction, by multiplying the interferogram by a complex exponential defined in Eq. (s1) prior to the FFT. When the comb spectrum becomes broad, the assumption of a uniform frequency shift across all comb modes no longer holds, and the resulting mismatch increases. In such cases, a zero-padding procedure is required to suppress this residual mismatch. In this work, with the spectral width of FWHM of 80 cm$^{-1}$, the residual error by assuming uniform shift is negligibly small.

Under ideal conditions—namely, when the system is operated in vacuum and the He-Ne laser and the frequency comb are perfectly co-aligned in the FTS—an ILS-free spectrum can be obtained simply by accurately determining the He-Ne laser wavelength and applying the corresponding sampling-mismatch correction. In practice, our home-made FTS system operated in air experience additional errors due to the wavelength-dependent refractive index and small differences in the beam paths of the comb and the He–Ne reference laser. The actual sampling interval for the mid-IR comb can be expressed as follows,

$$x_{sampling} = \lambda_{HeNe,eff}/4 = \frac{n_{midIR}}{n_{633\ nm}}\lambda_{HeNe}(1+\eta)/4 \quad \text{(s4)}$$

where $\lambda_{\text{HeNe}}$ is the wavelength of the reference laser, $n_{633}$ and $n_{midIR}$ are the refractive indices at the corresponding wavelengths, and $\eta$ represents the wavelength error arising from imperfect beam alignment. We denote the He–Ne wavelength that includes both the refractive-index effects and the beam-alignment offset as effective He-Ne laser wavelength ($\lambda_{\mathrm{HeNe,eff}}$). By precisely determining the $\lambda_{\mathrm{HeNe,eff}}$, we can obtain an ILS-free spectrum while simultaneously assigning an accurate frequency axis to the resulting spectra. In this work, in order to achieve both ILS-free high-precision line-shape measurements and accurate absolute frequency calibration without relying on external references, we numerically quantified the ILS observed in the absorption spectra by varying the $\lambda_{\mathrm{HeNe,eff}}$ in small steps and evaluating the quality factor of the line-shape fit (QF) that is absorption depth over standard deviation of the fitting residual. For each absorption line in each measurement, the value of the $\lambda_{\mathrm{HeNe,eff}}$ that yields QF peak i.e. minimum ILS was calculated.

Figure s1(a) shows the QF of the R7 line measured at 10 Torr as a function of the $\lambda_{\mathrm{HeNe,eff}}$. The horizontal axis corresponds to the $\lambda_{\mathrm{HeNe,eff}}$ in units of MHz. When the $\lambda_{\mathrm{HeNe,eff}}$ wavelength is varied 400 MHz, the absolute frequency of the R7 line changes over a range of 55 MHz, as indicated on the upper axis. Figures s1(b) and s1(c) show the line shape of the R7 transition and its speed-dependent Voigt profile (SDVP) fit, together with the residuals, for the points giving the best and worst QF within the present scan range, respectively. The QF dependence is symmetric to the peak, and the $\lambda_{\mathrm{HeNe,eff}}$ corresponding to the QF peak was derived from the fit to even-order polynomial function. The QF exhibits periodic maxima at intervals corresponding to the repetition frequency $f_{\mathrm{rep}}$. Therefore, in order to determine the $\lambda_{\mathrm{HeNe,eff}}$, the absolute frequency of the measured spectral line must be known with an accuracy better than $f_{\mathrm{rep}}$. Once the $\lambda_{\mathrm{HeNe,eff}}$ has been determined, however, this value becomes an instrument-specific constant and does not vary significantly over time. Consequently, all subsequent measurements can be performed as fully reference-free and independent absolute frequency measurements. This approach has previously been used for the determination of the absolute frequency of sub-Doppler spectral lines in Ref. [19].

As expressed in Eq. (s4), the $\lambda_{\mathrm{HeNe,eff}}$ depends on the refractive index of air that depends on the optical wavelength. Figure s2 shows the $\lambda_{\mathrm{HeNe,eff}}$ that yield the peak QF for each transition in the R branch. The horizontal axis represents the transition frequency of each line. The error bars in the plot represent the uncertainty in determining the QF peak from the polynomial fit. We introduced a refractive-index term $n_{midIR}(\lambda)$, expressed using the mather-function formalism,

$$\lambda_{\mathrm{HeNe,eff}}(\lambda) = \frac{n_{midIR}(\lambda)}{n_{633\,nm}} \lambda_{HeNe}(1+\eta) \quad . \qquad \text{(s5)}$$

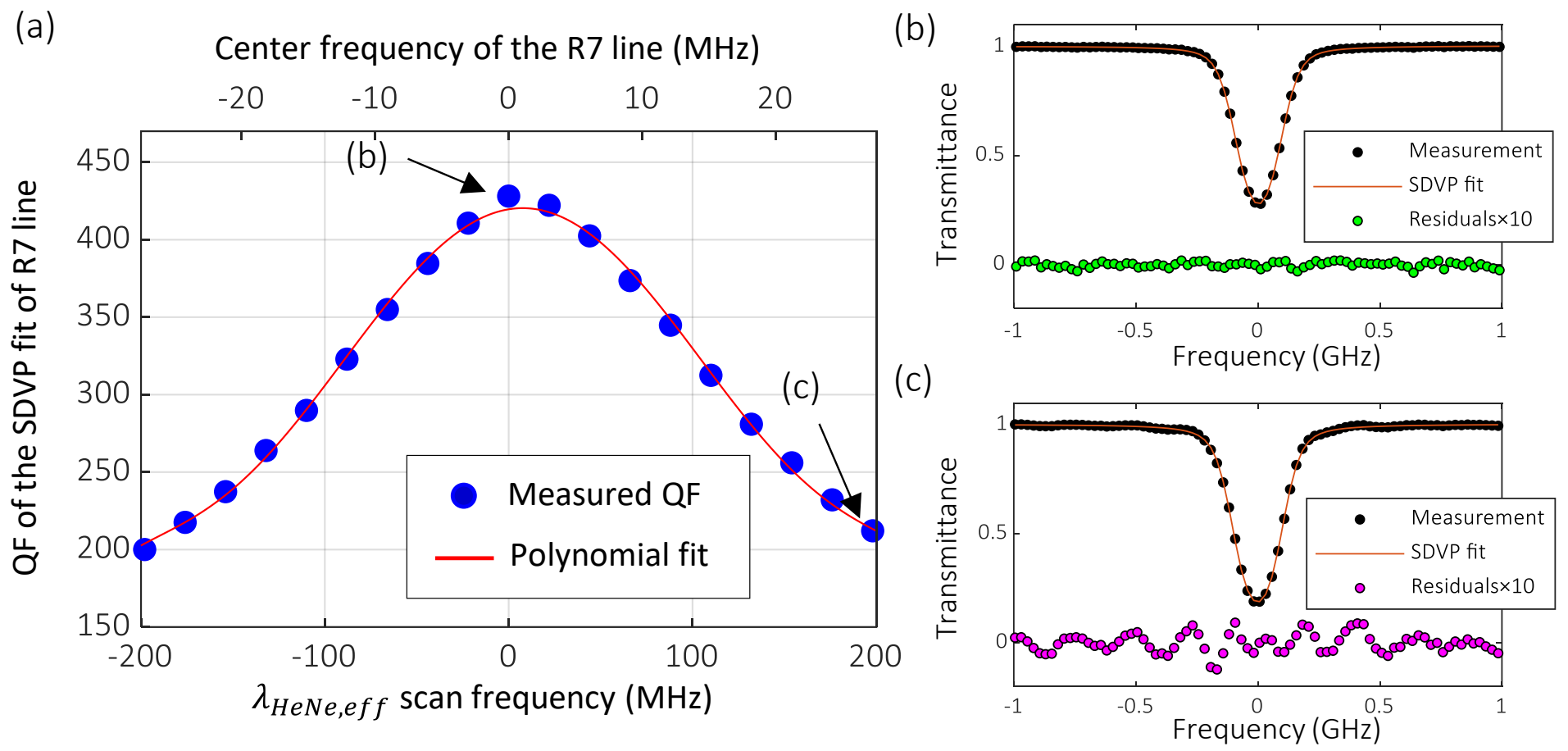

Fig. 1s. (a) Quality factor (QF) of the R7 transition measured at 10 Torr as a function of the effective He–Ne reference wavelength, $\lambda_{\mathrm{HeNe,eff}}$. The lower horizontal axis shows$\lambda_{\mathrm{HeNe,eff}}$ expressed in frequency units (MHz), while the upper axis indicates the corresponding shift in the absolute line-center frequency. (b, c) Measured line shape of the R7 transition and the corresponding speed-dependent Voigt profile (SDVP) fit, together with fit residuals, for the values of $\lambda_{\mathrm{HeNe,eff}}$ yielding the maximum (b) and minimum (c) QF within the scanned range. The residuals are magnified by a factor of 10 for clarity.

The red line in Fig. 2s represents the result of the fit using Eq. (5). Error bars were used as weights in the fit. Fitted value of $\lambda_{\mathrm{HeNe,eff}}(\lambda)$ was then used to determine the absolute center frequency of the spectra.

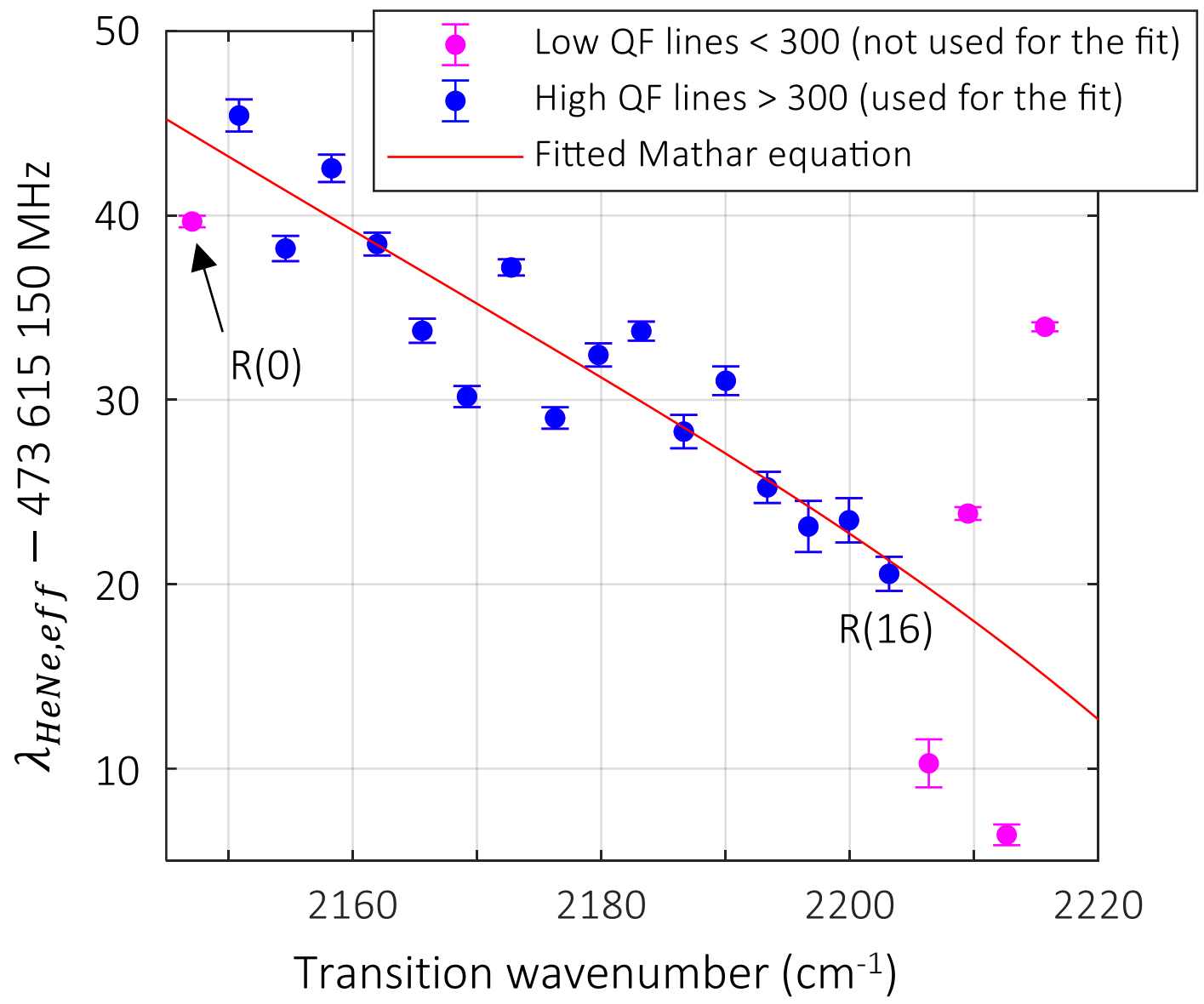

Fig. s2. The $\lambda_{\mathrm{HeNe,eff}}$ corresponding to the peak quality factor (QF) for individual transitions in the R branch, plotted as a function of the transition frequency. Error bars indicate the uncertainty in determining the QF peak from the even-order polynomial fit. Blue circles denote transitions with QF peak > 300 that were included in the fit using the mather-function–based refractive-index model, whereas pink circles indicate transitions with lower QF that were excluded from the fit. The red solid line represents the result of the fit based on Eq. (S5).

### *Amplitude correction*

We applied an interferogram flatness correction to suppress residual instrumental line-shape (ILS) effects. Owing to the long interferogram length (2.4 m corresponding to $c/f_{\mathrm{rep}}$), the interferogram amplitude was not uniform, effectively acting as a window function. A reference interferogram measured with a near-infrared cw laser was used to derive an amplitude correction curve, which was applied to the mid-IR comb interferograms. This correction alters the line shape by about 0.5% at line center for the 10-Torr R7 transition, resulting in a ~4% change in the fitted Lorentzian width. At higher pressures, the impact becomes negligible (<0.01% at 400 Torr).

### *Tables of the line position and collisional line parameters*

**Table s1. Zero-pressure line positions of the fundamental band of CO**

| P branch | | R branch | |
|---|---|---|---|
| *m* | $cm^{-1}$ | *m* | $cm^{-1}$ |
| -1 | 2139.426048 (6) | 1 | 2147.081165 (7) |
| -2 | 2135.546168 (4) | 2 | 2150.856022 (4) |
| -3 | 2131.631564 (3) | 3 | 2154.595606 (3) |
| -4 | 2127.682417 (5) | 4 | 2158.299724 (3) |
| -5 | 2123.698807 (3) | 5 | 2161.968259 (3) |
| -6 | 2119.680949 (2) | 6 | 2165.601055 (2) |
| -7 | 2115.628966 (2) | 7 | 2169.197962 (3) |
| -8 | 2111.543010 (2) | 8 | 2172.758837 (3) |
| -9 | 2107.423220 (2) | 9 | 2176.283531 (2) |
| -10 | 2103.269747 (3) | 10 | 2179.771894 (2) |
| -11 | 2099.082737 (2) | 11 | 2183.223792 (3) |
| -12 | 2094.862337 (2) | 12 | 2186.639062 (2) |
| -13 | 2090.608690 (3) | 13 | 2190.017568 (3) |
| -14 | 2086.321959 (3) | 14 | 2193.359164 (3) |
| -15 | 2082.002270 (4) | 15 | 2196.663695 (3) |
| -16 | 2077.649776 (4) | 16 | 2199.931023 (3) |
| -17 | 2073.264631 (5) | 17 | 2203.160997 (4) |
| -18 | 2068.846969 (7) | 18 | 2206.353484 (4) |
| -19 | 2064.396959 (9) | 19 | 2209.508325 (6) |
| -20 | 2059.914715 (13) | 20 | 2212.625362 (8) |

**Table s2. Collisional line-shape parameters**

| P branch | | | | R branch | | | |
|---|---|---|---|---|---|---|---|
| $m$ | $\delta_0$ cm$^{-1}$/atm | $\gamma_0$ cm$^{-1}$/atm | $a$ | $m$ | $\delta_0$ cm$^{-1}$/atm | $\gamma_0$ cm$^{-1}$/atm | $a$ |
| -1 | -0.00130 (18) | 0.07037 (10) | 0.0830 (30) | 1 | -0.00142 (17) | 0.07119 (11) | 0.0927 (32) |
| -2 | -0.00162 (9) | 0.06540 (5) | 0.0977 (16) | 2 | -0.00149 (9) | 0.06525 (5) | 0.0953 (16) |
| -3 | -0.00241 (8) | 0.06109 (4) | 0.1144 (14) | 3 | -0.00139 (7) | 0.06096 (4) | 0.1179 (13) |
| -4 | -0.00347 (8) | 0.05718 (3) | 0.1243 (12) | 4 | -0.00130 (6) | 0.05683 (3) | 0.1283 (12) |
| -5 | -0.00328 (7) | 0.05389 (4) | 0.1357 (15) | 5 | -0.00134 (6) | 0.05386 (3) | 0.1356 (12) |
| -6 | -0.00346 (7) | 0.05101 (3) | 0.1226 (12) | 6 | -0.00142 (6) | 0.05092 (2) | 0.1279 (11) |
| -7 | -0.00351 (6) | 0.04921 (3) | 0.1210 (12) | 7 | -0.00146 (6) | 0.04914 (2) | 0.1248 (11) |
| -8 | -0.00351 (6) | 0.04799 (2) | 0.1192 (12) | 8 | -0.00151 (6) | 0.04785 (2) | 0.1220 (11) |
| -9 | -0.00353 (6) | 0.04688 (2) | 0.1159 (12) | 9 | -0.00157 (6) | 0.04675 (2) | 0.1169 (11) |
| -10 | -0.00344 (7) | 0.04663 (3) | 0.1295 (12) | 10 | -0.00162 (6) | 0.0460 (2) | 0.1198 (11) |
| -11 | -0.00348 (6) | 0.04554 (3) | 0.1158 (13) | 11 | -0.00169 (6) | 0.04536 (2) | 0.1165 (11) |
| -12 | -0.00343 (7) | 0.04499 (3) | 0.1198 (13) | 12 | -0.00170 (6) | 0.04487 (2) | 0.1176 (11) |
| -13 | -0.00343 (7) | 0.04466 (3) | 0.1236 (15) | 13 | -0.00176 (6) | 0.04447 (2) | 0.1224 (11) |
| -14 | -0.00338 (7) | 0.04440 (4) | 0.1335 (16) | 14 | -0.00180 (6) | 0.04399 (2) | 0.1261 (12) |
| -15 | -0.00334 (8) | 0.04380 (4) | 0.1336 (18) | 15 | -0.00186 (6) | 0.04362 (3) | 0.1301 (13) |
| -16 | -0.00337 (9) | 0.04352 (6) | 0.1413 (23) | 16 | -0.00192 (6) | 0.04333 (3) | 0.1467 (14) |
| -17 | -0.00348 (10) | 0.04306 (7) | 0.1525 (28) | 17 | -0.00206 (7) | 0.04248 (4) | 0.1302 (17) |
| -18 | -0.00318 (15) | 0.04262 (10) | 0.1575 (41) | 18 | -0.00215 (8) | 0.04215 (5) | 0.1437 (20) |
| -19 | -0.00323 (21) | 0.04201 (15) | 0.1664 (59) | 19 | -0.00229 (10) | 0.04164 (7) | 0.1495 (28) |
| -20 | -0.00331 (28) | 0.04179 (23) | 0.1849 (84) | 20 | -0.00228 (12) | 0.04099 (9) | 0.1581 (38) |